\documentclass[11pt]{article}
\usepackage{longtable}
\usepackage{lscape}

\usepackage{amsmath}

\usepackage{amssymb}

\begin{document}




\title{{Table of hyperfine anomaly in atomic systems - 2023}}

  \author{J.R. Persson\\
  Department of Physics\\
  NTNU\\
  NO-7491 Trondheim\\
  Norway\\
  E-mail: jonas.persson@ntnu.no}



\date{24.04.2023} 
\maketitle

\begin{abstract}
 This table is an updated compilation of experimental values of the magnetic hyperfine anomaly in atomic and ionic systems. The literature search covers the period up to December 2022. A short discussion on general trends of the hyperfine anomaly and the theoretical developments are performed.
\end{abstract}





\newpage



\section{Introduction}
The effect on the hyperfine structure by nuclear charge and magnetisation distributions has been studied since the first experimental observation in Rb was made in 1949 by Bitter \cite{Bitter}. The effect of an extended nuclear magnetisation is manifested by the difference between the assumed point-like and actual magnetic dipole hyperfine interaction, as first anticipated by Kopfermann \cite{Kopf} towards the end of the 1930ties. The effect of extended magnetisation was calculated by Bohr and Weisskopf \cite{bohrweisskopf} in 1950 and is therefore known as the Bohr-Weisskopf (BW) effect. The extended nuclear charge distribution will, in addition to the Isotope Shift (IS), also give rise to an effect on the hyperfine structure. This is due to a modification of the electronic wavefunctions by the extended nuclear charge distribution, the so-called "Breit-Rosenthal-Crawford-Schawlow", or Breit-Rosenthal(BR) correction \cite{Breit,Crawford,Pallas,Rosenberg}.

These effects are normally quite small in atomic systems and remain quite difficult to calculate accurately, except in one-electron systems. This is the reason why it is more sensible to to study differential differences between isotopes, the so-called differential hyperfine anomaly. Due to the small size of the differential effect only a few systematic investigations have been performed \cite{stroke,mosko,Buttgenbach, PERSSON201362}. This review is an update of the 2013 paper as there have been some developments since it was published.

\section{Hyperfine Anomaly} 

The only way to probe the structural properties of the nucleus is with the use of "penetrating" electrons, that is the electrons wavefunction must have a nonzero probability to be found inside the nucleus. This means that only $s_{1/2}$ and relativistic $p_{1/2}$-electrons can be used. One should note that electron-electron interaction can cause an s- or p- electron-like behaviour, thus giving rise to an apparent nonzero probability for other electrons. A more pronounced effect can be obtained with the use of muons, this is outside the scope of this review.

The magnetic dipole hyperfine interaction is represented by the Hamiltonian
	\begin{equation}
    H=aI \cdot J,
    \end{equation}
Where $a$ is the magnetic dipole hyperfine interaction constant. I and J are the nuclear and electron angular momenta, respectively. In the case of an extended nucleus, the point-like hyperfine interaction constant $a_{point}$, will be modified by two effects:\\
1.	The modification of the electronic wavefunctions by the extended nuclear charge distribution, the "Breit-Rosenthal-Crawford-Schawlow" correction ($\epsilon_{BR}$) \cite{Breit,Crawford,Pallas,Rosenberg}.\\
2.	The extended and distributed nuclear magnetisation, the Bohr-Weisskopf effect,($\epsilon_{BW}$ )\cite{bohrweisskopf}.\\
This yields
    \begin{equation}
    a=a_{point}\left(  1+\epsilon_{BW}\right)\left(  1+\epsilon_{BR}\right)
    \end{equation}
Where $a$ denotes the experimental value of the magnetic dipole hyperfine interaction constant.
The hypothetical $a_{point}$ can normally not be calculated with sufficient precision for ordinary atoms, as is the case in muonic atoms and hydrogen-like ions. However, these uncertainties in point-like interactions cancel if we take the ratio of the $a$ values for two isotopes. That is, we can express the ratio without the need for precise calculations:
	\begin{equation}
    \frac{a_{1}}{a_{2}}=
    {\frac{g_I{(1)}}{g_I{(2)}}
    {\frac{[1+\epsilon_{BW}(1)][1+\epsilon_{BR}(1)]}{[1+\epsilon_{BW}(2)][1+\epsilon_{BR}(2)]}}}
    \end{equation}

Where ${g_I = -\mu_I/I}$ is the nuclear gyromagnetic ratio. As we can assume, in the case of electrons, that $\epsilon$  are generally small, we can rewrite (3) as:
    \begin{eqnarray}
    \frac{a_{1}}{a_{2}}\approx{\frac{g_I{(1)}}{g_I{(2)}}{[1+\epsilon_{BW}(1)-\epsilon_{BW}(2)][1+\epsilon_{BR}(1)-\epsilon_{BR}(2)]}}\nonumber\\ ={\frac{g_I{(1)}}{g_I{(2)}}{[1+^{1}\Delta^{2}_{BW}][1+^{1}\Delta^{2}_{BR}]}}
    \end{eqnarray}
	
Here we define the differential hyperfine anomaly of Bohr-Weisskopf and Breit-Rosenthal corrections, respectively as:
    \begin{equation}
    ^{1}\Delta^{2}_{BW}\equiv \epsilon_{BW}(1)-\epsilon_{BW}(2)
    \end{equation}
    \begin{equation}
    ^{1}\Delta^{2}_{BR}\equiv \epsilon_{BR}(1)-\epsilon_{BR}(2)
    \end{equation}
	
Calculations of $\epsilon_{BR} $ may show significant values, while the differential $^{1}\Delta^{2}_{BR}$ is normally expected to be small and negligible compared to $^{1}\Delta^{2}_{BW}$\cite{Rosenberg}. However, in cases where the nuclei are similar with respect to nuclear spin and nuclear configuration $^{1}\Delta^{2}_{BR}$ can be on the same order as $^{1}\Delta^{2}_{BW}$ \cite{Heggset2020} and should be calculated. 

Since the BR effect can be neglected or calculated the subscripts can be dropped obtaining
    \begin{equation}
    \frac{a_{1}}{a_{2}}\approx{\frac{g_I{(1)}}{g_I{(2)}}{[1+^{1}\Delta^{2}]}}
    \end{equation}
	
In most cases will the hyperfine anomaly be of the order of $10^{-3}$ , making it  necessary to know the hyperfine interaction constants, $a$, and the nuclear gyromagnetic values ($\frac{\mu_I}{I}$) with at least an accuracy of $10^{-4}$ or better to obtain values accurate to 10\% for the hyperfine anomaly\cite{Buttgenbach}. Precision values of the hyperfine interaction constants, $a$, and independently measured nuclear magnetic dipole moments ($\mu_I)$ are thus needed to obtain the differential hyperfine anomaly,$^{1}\Delta^{2}$.

\section{State-dependent and s-electron hyperfine anomaly}
It has been observed that the hyperfine anomaly is depending on the atomic state, where the values for different states can vary significantly, but generally shows an n-independence, as found in Rb \cite{Galvan}. It should be noted that the n-independence is the case in single s- and p-electron states.  While the hyperfine anomalies normally are on the order of 1\% or less, the state-dependent hyperfine anomaly can attain values up to 10 \% in special cases. The origin of this can be found in the hyperfine interaction operators and different contributions to the hyperfine interaction.

The hyperfine interaction can be represented by the following operators\cite{Ref4,Ref5}:
\begin{equation}
h = \frac{{\mu _0 }}{{4\pi }}2\mu _B \sum\limits_{i = 1}^N {\left[ {{\mathop{\rm l}\nolimits} \left\langle {r^{ - 3} } \right\rangle ^{01}  - \sqrt {10} \left( {{\mathop{\rm sC^2}\nolimits}  } \right)^1 \left\langle {r^{ - 3} } \right\rangle ^{12}  + {\mathop{\rm s}\nolimits} \left\langle {r^{ - 3} } \right\rangle ^{10} } \right]} _i  \cdot \mu _I,
\end{equation}
where ${{\mathop{\rm l}\nolimits}}$ and ${\mathop{\rm s}\nolimits}$ are the orbital and spin angular momentum operators, respectively, of the electron, ${{\mathop{\rm sC^2}\nolimits}  }$ is a tensor product of ${\mathop{\rm s}\nolimits}$ and ${\mathop{\rm C^2}\nolimits}$ of rank 1. The indices stand for the rank in the spin and orbital spaces, respectively. The hyperfine interaction can be divided into three parts, orbital, spin-dipole and contact (spin) interaction, where only the contact (spin) interaction contributes to the hyperfine anomaly. This means that only s and $p_{1/2}$ electrons contribute to the hyperfine anomaly through the contact (spin) interaction. We can thus write the $a$ constant as consisting of a contact and non-contact part:
\begin{equation}
a=a_{nc}+a_{c},
\end{equation}

Where $a_{c}$ is the contribution due to the contact interaction and $a_{nc}$ the contribution due to non-contact interactions. The experimental hyperfine anomaly, defined with the experimental $a$ constant, should then be rewritten to obtain the contact contribution to the hyperfine anomaly:
\begin{equation}
{^{1}\Delta_{exp}^{2}}={^{1}\Delta_{c}^{2}}{\frac{a_{c}}{a}}
\end{equation}
where $^{1}\Delta^{2}_{c}$ is the hyperfine anomaly due to the contact interaction, that is, for an $s$- or $p_{1/2}$-electron. The hyperfine anomaly is most often given as the experimental state-dependent hyperfine anomaly, as the contact (s-electron) anomaly can be difficult to extract in the case of many-electron configurations.

However, observing that the contact anomaly $^{1}\Delta_{c}^{2}$ is a property that is independent of the state and constant within the atom. One finds that it is possible to extract the anomaly solely from the $a$-constants of two different atomic states \cite{persson}, provided the ratio $a_c/a$ differs substantially for the states. Comparing the ratio of $a$-constants for two isotopes in two atomic levels then gives:
\begin{equation}
{\frac{{a_{B}^{(1)}/a_{B}^{(2)}}}{{a_{C}^{(1)}/a_{C}^{(2)}}}}\approx
{1+{^{1}\Delta_{c}^{2}}({{\frac{a_{c}^{B}}{a^{B}}}-{\frac{a_{c}^{C}}{{a^{C}}}
})}}
\end{equation}
Where B and C denote different atomic levels and 1 and 2 denote different isotopes. The ratio between the two $a$-constant ratios ($\frac{a_{c}^{X}}{a^{X}}$) for the isotopes will only depend on the  difference in the contact contributions of the two atomic levels and the hyperfine anomaly for the s-electron and will be the same for all isotopes. It should be pointed out that the atomic states used must differ significantly in the ratio $\frac{a_{c}}{a}$, as a small difference will lead to an increased sensitivity to errors \cite{persson}. 
If we have a state where the hyperfine anomaly is expected to be small, the ${\frac{a_{c}^{C}}{{a^{C}}}}$ can be set to zero for that state, thus serving as a substitute for $g_I$. It is then possible to obtain values of the contact hyperfine anomaly for level B. 
Since the contact interaction is proportional to the $g_J$-factor of the state, it is easy to identify suitable candidates from the values of the theoretical or experimental $g_J$-factor . The optimal case would be two atomic levels within the same multiplet where the experimental (or theoretical) $g_J$ is greater and smaller than 1, respectively.
This method is especially useful for unstable isotopes where high-precision measurements of the nuclear magnetic moment do not exist. 
Furthermore, states with a substantial difference in  the ratio $a_c/a$ are also preferable for studies of the isotope shift. If measurements are performed on more than three atomic levels, it is also possible to deduce the nuclear magnetic moment ratio and the hyperfine anomaly simultaneously. This will give the nuclear magnetic moment with high accuracy provided the nuclear magnetic moment is known for at least one stable isotope.
The ratio $a_c/a$ can also be calculated with high accuracy making this method very useful where calculations can be done. 

\section{Calculation related to hyperfine anomaly}
Calculations of the BW-effect and subsequently hfa have long been performed by modelling the nuclear magnetisation using a uniform or Fermi distribution of magnetisation. This combined with difficulties with the accuracy of wavefunctions close and within the nucleus for many-electron systems, has given values far from experimental values. However, progress, both theoretically and experimentally, has made it possible to link few-electron (highly charged ions) and muonic systems to calculations of many-electron systems \cite{Sanamyan}. This development has been a by-product of the effort for finding new physics in precision atomic searches \cite{Sanamyan}.  
The work of Sanamyan et al. \cite{Sanamyan} also showed that the simplistic model of uniform magnetisation do not describe the BW-effect in a good way and should be replaced with other available models such as different  varieties of the single-particle model.
Simultaneously, work on improving the GRASP code is on the way \cite{Jonsson} including the possibility to calculate the BW-effect. This will open possibilities for a systematic studies as it will become easier to compare experimental values with own calculations. In addition, as was found by Heggset et. al\cite{Heggset2020} the BR-effect can be calculated readily using the present GRASP code, it is still interesting to study the effect using a more realistic charge distribution for heavy nuclei \cite{Papoulia}.

\section{Trends in hyperfine anomaly and need for more measurements}
Even if the hfa has been studied since the 1950ties, the values for different elements are sporadic and very few systematic studies have been done due to the requirements of high-precision measurements of both hyperfine structure and nuclear magnetic dipole moments. It is well established that the hfa (BW-effect) attains rather similar values within the same nuclear configuration and spin as has been observed in Fr (I=9/2), Hg (I=13/2), Pb (I=13/2) and Eu (I=5/2). In these cases, the BR-effect may be on the same order as the BW-effect \cite{Heggset2020} and can not be neglected. Since it is possible to calculate the BR-effect it is possible to assume a smaller uncertainty in the determination of the nuclear magnetic dipole moment of unstable isotopes if the hfa (BW- and BR-effect) is known in an isotope with the same nuclear configuration. This is of special importance in laser spectroscopy studies in elements with $Z<83$.
The procedure to use the Moscowitz-Lombardi semi-empirical formula as a tool to estimate the hfa is questionable, as it has been shown that it is not universal and only seems to be valid in the Z=80 region. However this should be studied further with more systematic measurements in this region.
There is also a need for more systematic studies in the vicinity of the alkali elements as these are used in precision atomic calculations with the aim to find new physics \cite{Sanamyan}.

\section{Outlook}
Even if the hyperfine anomaly and BW-effect has been a small subject, progress both experimentally and in calculations open the possibility for more systematic studies with better understanding of nuclear structure as well as a means to test the atomic wavefunctions in precision atomic calculations. From that perspective one can expect a renewed interest and more precision measurements of the hyperfine structure and from these, the hyperfine anomaly.

\section{Policies followed in the compilation}

The hyperfine anomaly is given with the lightest stable isotope as the reference isotope. The lightest naturally abundant isotope was used for U and the designation of the original article was used for Fr. In most cases the original article, where the hyperfine anomaly has been derived, is used. In the case of recent, more precise values of the nuclear magnetic moment, the hyperfine anomaly has been updated, accordingly. The nuclear magnetic moments of Stone \cite{stone} have been used, unless more precise values of ratios are available. Special care was taken to use magnetic moments obtained by the same method.
The hyperfine anomaly is given as state-dependent if not stated otherwise. If the s-electron hyperfine anomaly is known, no extensive listing of state-dependent hyperfine anomaly is given, unless these are of special interest.




\clearpage

\section*{Table 1.\label{tbl1te} Experimental data of hyperfine anomaly values in atomic systems}

\begin{center}
\begin{tabular}{ll}

Element & The element studied\\

Isotope 1
	& Reference isotope for the hyperfine anomaly\\
Spin 1
	& Spin for isotope 1\\

Isotope 2
	& The second isotope used.\\

Spin 2
	& Spin for isotope 2\\

State-/s- anomaly	& The atomic state for which the experimental hfa\\
                        & has been determined or the s-electron hfa.\\

$^{1}\Delta^{2}(\%)$
	& Hyperfine anomaly given in \%.\\

Reference
	& Original article where $^{1}\Delta^{2}(\%)$ or the experimental hyperfine\\
    & interactions constants is given.\\

\end{tabular}
\end{center}
\label{tableI}


\bigskip

\newpage

\begin{landscape}




\setlength{\LTleft}{0pt}
\setlength{\LTright}{0pt}


\setlength{\tabcolsep}{0.5\tabcolsep}

\renewcommand{\arraystretch}{1.0}

\footnotesize 

\begin{longtable}{lcccclcr}
\caption{Experimental data of hyperfine anomaly values in atomic systems}\\
\mbox{Element} & \mbox{isotope 1} & \mbox{spin 1}& \mbox{isotope 2} & \mbox{spin 2}& Atomic state/ s-anomaly & $^{1}\Delta^{2}(\%)$ & Reference\\
\hline\\
\endfirsthead\\
\caption[]{(continued)}\\
\mbox{Element} & \mbox{isotope 1} & \mbox{spin 1}& \mbox{isotope 2} & \mbox{spin 2}& Atomic state/ s-anomaly & $^{1}\Delta^{2}(\%)$ & Reference\\
\hline\\
\endhead

Li & 6 & 1 & 7 & 3/2 & 2s $^{2}$S$_{1/2}$, s-anomaly & 0.0068067(8) & \cite{Allegrini22} \\ 
\\
&  &&&  & 3s $^{2}$S$_{1/2}$, s-anomaly & 0.054(21) & \cite{Allegrini22} \\
&  &&&  & 2p $^{2}$P$_{1/2}$ & -0.1734(2) & \cite{Allegrini22} \\
&  &&&  & 2p $^{2}$P$_{3/2}$ & -0.155(8) & \cite{Allegrini22} \\
\\

N & 14 & 1 & 15 & 1/2 & 2p$^{3}$ $^{4}$S$_{3/2}$ & 0.0999(4) & \cite{Holloway1962} \\
\\

Na & 23 & 3/2 & 24 & 4 & 3s $^{2}$S$_{1/2}$, s-anomaly & 0.0013(30) & \cite{Beckmann1974}\cite{Chan1966} \\
\\

Mg & 25 & 5/2 & 23 & 3/2 & 3s $^{2}$S$_{1/2}$ - 3p $^{2}$P$_{1/2}$ ($\Delta _{s}-\Delta
_{p}$) & -0.04(8) & \cite{Yordanov2017}\\
\\

Cl & 35 & 3/2 & 37 & 3/2 & 3p$^{5}$ $^{2}$P$_{3/2}$ & -0.00381(2) & \cite{King1956}
\\
\\

K & 39 & 3/2 & 37 & 3/2 & 4s $^{2}$S$_{1/2}$, s-anomaly & -0.249(35) & \cite{Platen1971}
\\
& 39 & 3/2 & 40 & 4 & 4s $^{2}$S$_{1/2}$, s-anomaly & 0.467(2) & \cite{Allegrini22}
\\

& 39 & 3/2 & 41 & 3/2 & 4s $^{2}$S$_{1/2}$, s-anomaly & -0.22937(13) & \cite{Beckmann1974}\cite{Allegrini22} \\

& 39 & 3/2 & 42 & 2 & 4s $^{2}$S$_{1/2}$, s-anomaly & 0.336(38) & \cite{Chan1969} \\

 & 39 & 3/2 & 38 & 3 &  4s $^{2}$S$_{1/2}$ - 4p $^{2}$P$_{1/2}$ ($\Delta _{s}-\Delta
_{p}$) & 0.53(44) & \cite{Papuga2014}\\
 & 39 & 3/2 & 40 & 4 &  4s $^{2}$S$_{1/2}$ - 4p $^{2}$P$_{1/2}$ ($\Delta _{s}-\Delta
_{p}$) & 0.43(17) & \cite{Papuga2014}\\
 & 39 & 3/2 & 41 & 3/2 &  4s $^{2}$S$_{1/2}$ - 4p $^{2}$P$_{1/2}$ ($\Delta _{s}-\Delta
_{p}$) & -0.23(31) & \cite{Papuga2014}\\
 & 39 & 3/2 & 42 & 2 &  4s $^{2}$S$_{1/2}$ - 4p $^{2}$P$_{1/2}$ ($\Delta _{s}-\Delta
_{p}$) & 0.99(36) & \cite{Papuga2014}\\
 & 39 & 3/2 & 44 & 2 &  4s $^{2}$S$_{1/2}$ - 4p $^{2}$P$_{1/2}$ ($\Delta _{s}-\Delta
_{p}$) & 0.47(47) & \cite{Papuga2014}\\
 & 39 & 3/2 & 46 & 2 &  4s $^{2}$S$_{1/2}$ - 4p $^{2}$P$_{1/2}$ ($\Delta _{s}-\Delta
_{p}$) & 0.40(39) & \cite{Papuga2014}\\
 & 39 & 3/2 & 47 & 1/2 &  4s $^{2}$S$_{1/2}$ - 4p $^{2}$P$_{1/2}$ ($\Delta _{s}-\Delta_{p}$) & 0.28(16) & \cite{Papuga2014}\\
 & 39 & 3/2 & 48  & 1 &  4s $^{2}$S$_{1/2}$ - 4p $^{2}$P$_{1/2}$ ($\Delta _{s}-\Delta
_{p}$) & 0.57(35) & \cite{Papuga2014}\\
 & 39 & 3/2 & 49 & 1/2 &  4s $^{2}$S$_{1/2}$ - 4p $^{2}$P$_{1/2}$ ($\Delta _{s}-\Delta
_{p}$) & 0.24(29) & \cite{Papuga2014}\\
 & 39 & 3/2 & 51 & 3/2 &  4s $^{2}$S$_{1/2}$ - 4p $^{2}$P$_{1/2}$ ($\Delta _{s}-\Delta
_{p}$) & 0.57(250) & \cite{Papuga2014}\\

\\
V & 50 & 6 & 51 & 7/2 & 3d$^{3}$4s$^{2}$ $^{4}$F$_{5/2}$ & 0.0007(10) & \cite{LUTZ1981}
\\
&  &&&  & 3d$^{4}$4s $^{6}$D$_{1/2}$ & 0.034(60) & \cite{Cochrane1998} \\
\\

Cu & 63 & 3/2 & 65 & 3/2 & 3d$^{10}$4s $^{2}$S$_{1/2}$ & 0.004861(9) & \cite{LUTZ1978}
\\
&  &&&  & 3d$^{9}$4s4p $^{4}$P$_{5/2}$ & 0.00340(11) & \cite{LUTZ1978}\cite%
{BLACHMAN1969} \\
&  &&&  & 3d$^{9}$4s4p $^{4}$P$_{9/2}$ & 0.00305(17) & \cite{LUTZ1978}\cite%
{BLACHMAN1969} \\
\\

Ga & 69 & 3/2 & 67 & 3/2 & 4p $^{2}$P$_{1/2}$ & -0.00050(7) & \cite{LUTZ1971} \\
&  & & &  & 4p $^{2}$P$_{3/2}$ & 0.00200(16) & \cite{LUTZ1971} \\
& 69 & 3/2 & 71 & 3/2 & 4p $^{2}$P$_{1/2}$ & 0.00063(6) & \cite{LUTZ1971} \\
&  & & &  & 4p $^{2}$P$_{3/2}$ & -0.00252(12) & \cite{LUTZ1971} \\
& 71 & 3/2 & 72 & 3 & 4p $^{2}$P$_{1/2}$ & 0.0043(6) & \cite{LUTZ1971} \\
&  & & &   & 4p $^{2}$P$_{3/2}$ & -0.0170(18) & \cite{LUTZ1971} \\
\\

As & 75 & 3/2 & 70 & 4 & 4p$^{3}$ $^{4}$S$_{3/2}$ & -0.35(2) & \cite{Hogervorst1980} \\
\\
Br & 79 & 3/2 & 81 & 3/2 & 4p$^{5}$ $^{2}$P$_{3/2}$ & -0.00003(4) & \cite{BROWN1966}\cite%
{LUTZ1970} \\
\\
Rb & 85 & 5/2 & 84 & 2 & 5s $^{2}$S$_{1/2}$, s-anomaly & -1.7(1.0) & \cite%
{ACKERMANN1973} \\
& 85 & 5/2 & 86 & 2 & 5s $^{2}$S$_{1/2}$, s-anomaly & 0.17(9) & \cite{BRASLAU1961} \\
& 85 & 5/2 & 87 & 3/2 & 5s $^{2}$S$_{1/2}$, s-anomaly & 0.35141(2) & 
\cite{Allegrini22} \\
&  &&&  & 6s $^{2}$S$_{1/2}$, s-anomaly & 0.361(19) & \cite{Allegrini22}\cite{PerezGalvan2007}\cite{PerezGalvan2008} \\
&  &&&  & 7s $^{2}$S$_{1/2}$, s-anomaly & 0.342(3) & \cite{Allegrini22} \\
&  &&&  & 5p $^{2}$P$_{1/2}$ & 0.55(8) & \cite{Allegrini22}\\
&  &&&  & 5p $^{2}$P$_{3/2}$ & 0.168(5) & \cite{Allegrini22} \\
&  &&&  & 6p $^{2}$P$_{1/2}$ & 0.31(7) & \cite{Allegrini22} \\
&  &&&  & 6p $^{2}$P$_{3/2}$ & 0.46(5) & \cite{Allegrini22} \\

&  &&&  & 4d $^{2}$D$_{3/2}$ & 0.347(4) & \cite{moon2009} \\
&  &&&  & 4d $^{2}$D$_{5/2}$ & 0.41(9) & \cite{Wang2014} \\
&  &&&  & 4d $^{2}$D$_{5/2}$ & 0.60(15) & \cite{Allegrini22} \\
&  &&&  & 5d $^{2}$D$_{3/2}$ & 0.279(6) & \cite{Allegrini22} \\
&  &&&  & 5d $^{2}$D$_{5/2}$ & 0.44(5) &\cite{Allegrini22} \\

\\
Mo & 95 & 5/2 & 97 & 5/2 & 4d$^{5}$5s $^{7}$S$_{3}$ & -0.0101(2) & \cite{Buttgenbach1974}
\\
\\
Ru & 99 & 5/2 & 101 & 5/2 & s-anomaly & -0.0173(1) & \cite{Buttgenbach1977} \\
\\

Ag & 107 & 1/2 & 103 & 7/2 & 4d$^{10}$5s $^{2}$S$_{1/2}$ & -3.4(1.7) & \cite{Wannberg1970} \\
& 107 & 1/2 & 108 & 1 & 4d$^{10}$5s $^{2}$S$_{1/2}$ & -2.6(7) & \cite{CUSSENS1969} \\
& 107 & 1/2 & 109 & 1/2 & 4d$^{10}$5s $^{2}$S$_{1/2}$ & -0.41274(29) & \cite{DAHMEN1967}
\\
& 107 & 1/2 & 109$^{m}$ & 7/2 & 4d$^{10}$5s $^{2}$S$_{1/2}$ & -3.8(4.1) & \cite%
{STINSON1971} \\
&  &&&  &  & -0.85(1.19) & \cite{STINSON1971}, $\mu _{I}$ from \cite{EDER1985}
\\
& 107 & 1/2 & 110 & 1 & 4d$^{10}$5s $^{2}$S$_{1/2}$ & -3.1(1.4) & \cite{CUSSENS1969}
\\
& 107 & 1/2 & 110$^{m}$ & 6 & 4d$^{10}$5s $^{2}$S$_{1/2}$ & -2.88(13) & \cite{Schmelling1967}\\
\\

Cd & 111 & 1/2 & 107 & 1/2 & 5s $^{2}$S$_{1/2}$ & -0.14(3) & \cite{Yordanov2013} \\
& 111 & 1/2 & 109 & 5/2& 5s $^{2}$S$_{1/2}$ & -0.14(3) & \cite{Yordanov2013} \\
& 111 & 1/2 & 113 & 1/2 & 5s $^{2}$S$_{1/2}$ & -0.02(3) & \cite{Yordanov2013} \\
& 111 & 1/2 & 113$^{m}$ & 11/2& 5s $^{2}$S$_{1/2}$ & -0.13(3) & \cite{Yordanov2013} \\
& 111 & 1/2 & 115 & 1/2 & 5s $^{2}$S$_{1/2}$ & 0.02(3) & \cite{Yordanov2013} \\
& 111 & 1/2 & 115$^{m}$ & 11/2& 5s $^{2}$S$_{1/2}$ & -0.08(3) & \cite{Yordanov2013} \\
\\

& 111 & 1/2 & 107 & 1/2 & 5s6s $^{3}$S$_{1}$ & -0.18(4) & \cite{Frommgen2015} \\
& 111 & 1/2 & 109 & 5/2 & 5s6s $^{3}$S$_{1}$ & -0.12(1) & \cite{Frommgen2015} \\
& 111 & 1/2 & 111$^{m}$ & 11/2 & 5s6s $^{3}$S$_{1}$ & -0.10(4) & \cite{Frommgen2015} \\
& 111 & 1/2 & 113 & 1/2 & 5s6s $^{3}$S$_{1}$ & -0.01(1) & \cite{Frommgen2015} \\
& 111 & 1/2 & 113$^{m}$ & 11/2& 5s6s $^{3}$S$_{1}$ & -0.08(1) & \cite{Frommgen2015} \\
& 111 & 1/2 & 115 & 1/2 & 5s6s $^{3}$S$_{1}$ & 0.05(3) & \cite{Frommgen2015} \\
& 111 & 1/2 & 113$^{m}$ & 11/2& 5s6s $^{3}$S$_{1}$ & -0.09(2) & \cite{Frommgen2015} \\
\\
& 111 & 1/2 & 107 & 1/2 & 5s5p $^{3}$P$_{1}$ & -0.0958(8) & \cite{THADDEUS1963} \\
& 111 & 1/2 & 109 & 5/2 & 5s5p $^{3}$P$_{1}$ & -0.0912(7) & \cite{THADDEUS1963} \\
& 111 & 1/2 & 113 & 1/2 & 5s5p $^{3}$P$_{1}$ & -0.00023(40) & \cite{CHANEY1969} \\
& 111 & 1/2 & 113$^{m}$ & 11/2& 5s5p $^{3}$P$_{1}$ & -0.0773(5) & \cite{CHANEY1969} \\
& 111 & 1/2 & 115 & 1/2 & 5s5p $^{3}$P$_{1}$ & 0.244(65) & \cite{CHANEY1969} \\
& 111 & 1/2 & 115$^{m}$ & 11/2& 5s5p $^{3}$P$_{1}$ & -0.236(90) & \cite{CHANEY1969} \\
\\
& 111 & 1/2 & 107 & 1/2 & 5s5p $^{3}$P$_{2}$ & -0.17(5) & \cite{Frommgen2015} \\
& 111 & 1/2 & 109 & 5/2 & 5s5p $^{3}$P$_{2}$ & -0.12(1) & \cite{Frommgen2015} \\
& 111 & 1/2 & 111$^{m}$ & 11/2 & 5s5p $^{3}$P$_{2}$ & -0.08(4) & \cite{Frommgen2015} \\
& 111 & 1/2 & 113 & 1/2 & 5s5p $^{3}$P$_{2}$ & -0.00143(6) & \cite{FAUST1960} \\
& 111 & 1/2 & 113 & 1/2 & 5s5p $^{3}$P$_{2}$ & 0.00(1) & \cite{Frommgen2015} \\
& 111 & 1/2 & 113$^{m}$ & 11/2& 5s5p $^{3}$P$_{2}$ & -0.08(2) & \cite{Frommgen2015} \\
& 111 & 1/2 & 115 & 1/2 & 5s5p $^{3}$P$_{2}$ & 0.01(2) & \cite{Frommgen2015} \\
& 111 & 1/2 & 115$^{m}$ & 11/2 & 5s5p $^{3}$P$_{2}$ & -0.09(4) & \cite{Frommgen2015} \\
\\

In & 113 & 9/2 & 115 & 9/2 & 5p $^{2}$P$_{1/2}$ & 0.00075(13) & \cite{ECK1957} \\
&  &&&  & 5p $^{2}$P$_{3/2}$ & -0.00238(13) & \cite{ECK1957} \\
\\
Sn & 115 & 1/2 & 117 & 1/2 & 5p$^{2}$ $^{3}$P$_{1}$ & 0,0034(10) & \cite{CHILDS1965} \\
&  &&&  & 5p$^{2}$ $^{3}$P$_{2}$ & -0.0003(10) & \cite{CHILDS1965} \\
& 117 & 1/2 & 119 & 1/2 & 5p$^{2}$ $^{3}$P$_{1}$ & 0.0049(10) & \cite{CHILDS1965} \\
&  &&&  & 5p$^{2}$ $^{3}$P$_{2}$ & -0.0009(10) & \cite{CHILDS1965} \\
&  &&&  & 5p$^{2}$ $^{1}$D$_{2}$ & +0.0001(10) & \cite{CHILDS1971} \\
\\
Sb & 121 & 5/2 & 123 & 7/2 & 5p$^{3}$ $^{4}$S$_{3/2}$ & -0.323(9) & \cite{FERNANDO1960}
\\
\\

Xe & 129 & 1/2 & 131 & 3/2 & 6s $^{2}$S$_{1/2}$, s-anomaly & 0.0440(44) & \cite{Faust1961}
\\
\\

Cs & 133 & 7/2 & 131 & 5/2 & 5p$^{5}$6s $^{3}$P$_{2}$, s-anomaly & 0.45(5) & \cite{WORLEY1965}
\\
& 133 & 7/2 & 134 & 4 & 6s $^{2}$S$_{1/2}$, s-anomaly & 0.169(30) & \cite{STROKE1957} \\
& 133 & 7/2 & 134$^{m}$ & 8 & 6s $^{2}$S$_{1/2}$, s-anomaly & -1.38(3) & \cite%
{COHEN1962} \\
& 133 & 7/2 & 135 & 7/2 & 6s $^{2}$S$_{1/2}$, s-anomaly & 0.037(9) & \cite{STROKE1957} \\
& 133 & 7/2 & 137 & 7/2 & 6s $^{2}$S$_{1/2}$, s-anomaly & 0.0018(40) & \cite{STROKE1957}
\\
\\
Ba & 135 & 3/2 & 137 & 3/2 & 5d6s $^{3}$D$_{1}$ & -0.205(7) & \cite{GUSTAVSSON1979} \\
&  &&&  & 5d6s $^{3}$D$_{2}$ & -0.179(22) & \cite{GUSTAVSSON1979} \\
&  &&&  & 5d6s $^{3}$D$_{3}$ & -0.188(17) & \cite{GUSTAVSSON1979} \\
&  &&&  & 5d6s $^{1}$D$_{2}$ & -0.212(26) & \cite{SCHMELLING1974} \\
&  &&&  & Ba$^{+}$ 6s $^{2}$S$_{1/2}$, s-anomaly & -0.191(5) & \cite{Trapp2000}
\\
\\
La & 138 & 5 & 139 & 7/2 & 5d6p $^{3}$D$_{1}$ & -0.35(23) & \cite{Iimura2003}\cite%
{CHILDS1979} \\
\\

Nd & 143 & 7/2 & 145 & 7/2 & s-anomaly & 0.2034(63) & \cite{Persson2018GdNd} \\
\\
Eu & 151 & 5/2 & 145 & 5/2 & s-anomaly & -0.08(15) & \cite{Persson2020} \\
& 151 & 5/2 & 146 & 4 & s-anomaly & 0.12(50) & \cite{Persson2020} \\
& 151 & 5/2 & 147 & 5/2 & s-anomaly & -0.12(17) & \cite{Persson2020} \\
& 151 & 5/2 & 148 & 5 & s-anomaly & 0.08(31) & \cite{Persson2020} \\
& 151 & 5/2 & 149 & 5/2 & s-anomaly & -0.19(16) & \cite{Persson2020} \\
& 151 & 5/2 & 150 & 5 & s-anomaly & 0.08(28) & \cite{Persson2020} \\
& 151 & 5/2 & 152 & 3 & s-anomaly & 0.50(6) & \cite{Persson2020} \\
& 151 & 5/2 & 153 & 5/2 & s-anomaly & -0.64(3) & \cite{Brand1981}\\
\\
Gd & 155 & 3/2 & 157 & 3/2 & s-anomaly & 0.106(24) & \cite{Persson2018GdNd} \\
\\
Dy & 161 & 5/2 & 163 & 5/2 & 4f$^{10}$6s6p $^{5}$K$_{8}$ & 0.019(16) & \cite{CLARK1982}
\\
&  &&&  & 4f$^{10}$6s6p $^{5}$K$_{9}$ & 0.025(11) & \cite{CLARK1982} \\
&  &&&  & 4f$^{10}$6s6p $^{5}$I$_{8}$ & -0.116(19) & \cite{CLARK1982} \\
&  &&&  & 4f$^{10}$6s6p $^{5}$H$_{7}$ & -0.176(36) & \cite{CLARK1982} \\
\\
Yb & 171 & 1/2 & 173 & 5/2 & 6s6p $^{3}$P$_{1}$ & -0.386(5) & \cite{BUDICK1970} \\
   &     &     &     &     &                    & -0.3857(51)& \cite{Atkinson2019}\\
&  &&&  & 4f$^{13}$5d6s$^{2}$ $^{3}$P$_{1}$ & 0.066(22) & \cite{BUDICK1970} \\
&  &&&  & Yb$^{+}$ 6s $^{2}$S$_{1/2}$, s-anomaly & -0.425(5) & \cite{Fisk1997}%
\cite{Munch1987} \\
\\

Lu & 175 & 7/2 & 176 & 7 & 5d6s$^{2}$ $^{2}$D$_{3/2}$ & 0.02(15) & \cite{Brenner1985}
\\
&  &&&  & 5d6s$^{2}$ $^{2}$D$_{5/2}$ & 0.19(15) & \cite{Brenner1985} \\
&  &&&  & 5d6s6p $^{4}$P$_{1/2}$ & 0.40(24) & \cite{Witte2002} \\
&  &&&  & 5d6s6p $^{4}$P$_{3/2}$ & 1.62(25) & \cite{Witte2002} \\
&  &&&  & 5d6s6p $^{4}$P$_{5/2}$ & 0.0(27) & \cite{Witte2002} \\
&  &&&  & 5d6s6p $^{4}$F$_{3,5,7/2}$, s-anomaly & 0.48(8) & \cite{Kuhnert1983}%
, $\mu _{I}$ from \cite{Brenner1985} \\
&  &&&  & 6s$^{2}$8p $^{2}$P$_{1/2}$ & 1.84(90) & \cite{Witte2002} \\
&  &&&  & 6s$^{2}$8p $^{2}$P$_{3/2}$ & 0.55(22) & \cite{Witte2002} \\
& 175 & 5/2 & 177 & 7/2 & s-anomaly & -0.018(35) & \cite{Brenner1985} \\
& 176 & 7 & 176$^{m}$ & 1 & s-anomaly & 0.48(8) & \cite{Brenner1985} \\
\\

Re & 185 & 5/2 & 186 & 1 & 5d$^{5}$6s$^{2}$ $^{6}$S$_{5/2}$ & -1.36(17) & \cite{Armstrong1965}\cite{Buttgenbach1981} \\
& 185 & 5/2 & 187 & 5/2 & 5d$^{5}$6s$^{2}$ $^{6}$S$_{5/2}$ & 0.031(8) & \cite{Armstrong1965} \\
& 185 & 5/2 & 187 & 5/2 & s-anomaly & 0.027(5) & \cite{Burger1982} \\
& 185 & 5/2 & 188 & 1 & 5d$^{5}$6s$^{2}$ $^{6}$S$_{5/2}$ & -1.28(28) & \cite{Armstrong1965}\cite{Buttgenbach1981} \\
\\

Ir & 191 & 3/2 & 193 & 3/2 & s-anomaly & -0.64(7) & \cite{Buttgenbach1978} \\
\\
\\
Au & 197 & 3/2 & 177 & 11/2 & 6s $^{2}$S$_{1/2}$ - 6p $^{2}$p$_{1/2}$ ($\Delta _{s}-\Delta
_{p}$) & 7.7(8) & \cite{Barzakh2020}\\
& 197 & 3/2 & 177 & 11/2 & s-anomaly & 11.4(14) & \cite{Barzakh2020}\\
& 197 & 3/2 & 185 & 5/2 & s-anomaly & 9.4(30) & \cite{Barzakh2020}\\
& 197 & 3/2 & 186 & 3 & s-anomaly & 3.1(51) & \cite{Barzakh2020}\\
& 197 & 3/2 & 187 & 1/2 & s-anomaly & 12.7(84) & \cite{Barzakh2020}\\
& 197 & 3/2 & 189 & 1/2 & s-anomaly & 9.4(59) & \cite{Barzakh2020}\\

 & 197 & 3/2 & 189 & 11/2 & 6s $^{2}$S$_{1/2}$ - 6p $^{2}$p$_{1/2}$ ($\Delta _{s}-\Delta
_{p}$) & 6.0(10) & \cite{Barzakh2020}
\\
& 197 & 3/2 & 189 & 11/2 & s-anomaly & 8.6(16) & \cite{Barzakh2020}\\

& 197 & 3/2 & 191 & 3/2 & s-anomaly & -1.2(14) & \cite{Barzakh2020}\\
 & 197 & 3/2 & 191 & 11/2 & 6s $^{2}$S$_{1/2}$ - 6p $^{2}$p$_{1/2}$ ($\Delta _{s}-\Delta
_{p}$) & 7.9(8) & \cite{Barzakh2020}\\
& 197 & 3/2 & 191 & 11/2 & s-anomaly & 11.7(14) & \cite{Barzakh2020}\\

& 197 & 3/2 & 193 & 3/2 & s-anomaly & -0.5(11) & \cite{Barzakh2020}\\
 & 197 & 3/2 & 193 & 11/2 & 6s $^{2}$S$_{1/2}$ - 6p $^{2}$p$_{1/2}$ ($\Delta _{s}-\Delta
_{p}$) & 7.6(6) & \cite{Barzakh2020}\\
& 197 & 3/2 & 193 & 11/2 & s-anomaly & 11.2(11) & \cite{Barzakh2020}\\

& 197 & 3/2 & 194 & 1 & s-anomaly & 1.8(33) & \cite{Barzakh2020}\\
 & 197 & 3/2 & 195 & 11/2 & 6s $^{2}$S$_{1/2}$ - 6p $^{2}$p$_{1/2}$ ($\Delta _{s}-\Delta
_{p}$) & 7.5(8) & \cite{Barzakh2020}\\
& 197 & 3/2 & 195 & 11/2 & s-anomaly & 11.4(14) & \cite{Barzakh2020}\\

& 197 & 3/2 & 196 & 2& 6s $^{2}$S$_{1/2}$, s-anomaly & 8.69(26) & \cite{Schmelling1970} \cite{EKSTROM1980} \\
& 197 & 3/2 & 198 & 2 & 6s $^{2}$S$_{1/2}$, s-anomaly & 8.53(8) & \cite{Bout1967}
\cite{EKSTROM1980} \\
& 197 & 3/2 & 199 & 3/2 & 6s $^{2}$S$_{1/2}$, s-anomaly & 3.64(29) & \cite{Bout1967}
\cite{EKSTROM1980} \\
\\

Hg & 199 & 1/2 & 193 & 3/2 & 6s6p $^{3}$P$_{1}$ & -0.61(3) & \cite{Reimann1973} \\
& 199 & 1/2 & 193$^{m}$ & 13/2 & 6s6p $^{3}$P$_{1}$ & -1.0552(13) & \cite{Reimann1973} \\
& 199 & 1/2 & 195 & 1/2 & 6s6p $^{3}$P$_{1}$ & -0.1470(9) & \cite{Reimann1973} \\
& 199 & 1/2 & 195$^{m}$ & 113/2 & 6s6p $^{3}$P$_{1}$ & -1.038(3) & \cite{Reimann1973} \\
& 199 & 1/2 & 197 & 1/2 & 6s6p $^{3}$P$_{1}$ & -0.0778(7) & \cite{Reimann1973} \\
& 199 & 1/2 & 197$^{m}$ & 13/2 & 6s6p $^{3}$P$_{1}$ & -1.021(3) & \cite{Reimann1973} \\
& 199 & 1/2 & 199$^{m}$ & 13/2 & 6s6p $^{3}$P$_{1}$ & --0.960(9) & \cite{Reimann1973} \\
& 199 & 1/2 & 201 & 3/2 & 6s6p $^{3}$P$_{1}$ & -0.1467(6) & \cite{Reimann1973} \\
& 199 & 1/2 &&  & 6s6p $^{3}$P$_{2}$ & -0.15653(4) & \cite{Reimann1973} \\
& 199 & 1/2 &&  & Hg$^{+}$, 6s $^{2}$S$_{1/2}$, s-anomaly & -0.16257(5) & \cite{Burt2009} \\
& 199 & 1/2 & 203 & 5/2 & 6s6p $^{3}$P$_{1}$ & -0.796(16) & \cite{Reimann1973} \\
\\
Tl & 203 & 1/2 & 205 & 1/2 & 6p $^{2}$P$_{1/2}$ & 0.01035(15) & \cite{LURIO1956} \\
   &     &&&     & 6p $^{2}$P$_{3/2}$ & -0.16258(10) & \cite{GOULD1956} \\
   &     &&&     & 7s $^{2}$S$_{1/2}$ & 0.0294(81) & \cite{Chen2012} \\
   &     &&&     & 7p $^{2}$P$_{1/2}$ & 0.05(4) & \cite{Ranjit2014} \\

   & 205 & 1/2 & 187 & 9/2 & 6p $^{2}$P$_{1/2}$ - 7s $^{2}$s$_{1/2}$ ($\Delta _{p}-\Delta
_{s}$) & 2.02(97) & \cite{Barzakh2012}\\
   & 205 & 1/2 & 189 & 9/2 & 6p $^{2}$P$_{1/2}$ - 7s $^{2}$s$_{1/2}$ ($\Delta _{p}-\Delta
_{s}$) & 1.47(105) & \cite{Barzakh2012}\\
   & 205 & 1/2 & 191 & 9/2 & 6p $^{2}$P$_{1/2}$ - 7s $^{2}$s$_{1/2}$ ($\Delta _{p}-\Delta
_{s}$) & 1.56(60) & \cite{Barzakh2012}\\ 
   & 205 & 1/2 & 193 & 9/2 & 6p $^{2}$P$_{1/2}$ - 7s $^{2}$s$_{1/2}$ ($\Delta _{p}-\Delta
_{s}$) & 1.31(62) & \cite{Barzakh2012}\\

Pb & 207 & 1/2 & 191 & 13/2 & 6p7s $^{3}$P$_{1}$ - 6p$^2$ $^{1}$D$_{2}$  & -1.72(68) & \cite{Persson2018Pb} \\
   & 207 & 1/2 & 193 & 13/2 & 6p7s $^{3}$P$_{1}$ - 6p$^2$ $^{1}$D$_{2}$ & -1.86(58) & \cite{Persson2018Pb} \\
   & 207 & 1/2 & 195 & 13/2 & 6p7s $^{3}$P$_{1}$ - 6p$^2$ $^{1}$D$_{2}$ & -1.53(70) & \cite{Persson2018Pb} \\
   & 207 & 1/2 & 197$^{m}$ & 13/2 & 6p7s $^{3}$P$_{1}$ - 6p$^2$ $^{1}$D$_{2}$ & -1.68(123) & \cite{Persson2018Pb} \\

\\
Fr & 212 & 5 & 206$^{g}$ & 3 & 7s $^{2}$S$_{1/2}$ - 7p $^{2}$P$_{1/2}$ ($\Delta
_{s}-\Delta _{p}$) & 0.026(30) & \cite{Zhang2015} \\
& 212 & 5 & 206$^{m}$ & 7 & 7s $^{2}$S$_{1/2}$ - 7p $^{2}$P$_{1/2}$ ($\Delta
_{s}-\Delta _{p}$) & -0.058(27) & \cite{Zhang2015} \\
& 212 & 5 & 207 & 9/2 & 7s $^{2}$S$_{1/2}$ - 7p $^{2}$P$_{1/2}$ ($\Delta
_{s}-\Delta _{p}$) & -0.349(29) & \cite{Zhang2015} \\
& 212 & 5 & 208 & 7 & 7s $^{2}$S$_{1/2}$ - 7p $^{2}$P$_{1/2}$ ($\Delta
_{s}-\Delta _{p}$) & -0.014(46) & \cite{Zhang2015} \\
& 212 & 5 & 208 & 7 & 7s $^{2}$S$_{1/2}$ - 7p $^{2}$P$_{1/2}$ ($\Delta
_{s}-\Delta _{p}$) & 0.032(38) & \cite{Grossman1999} \\
& 212 & 5 & 209 & 9/2 & 7s $^{2}$S$_{1/2}$ - 7p $^{2}$P$_{1/2}$ ($\Delta
_{s}-\Delta _{p}$) & -0.368(29) & \cite{Zhang2015} \\
& 212 & 5 & 209 & 9/2 & 7s $^{2}$S$_{1/2}$ - 7p $^{2}$P$_{1/2}$ ($\Delta _{s}-\Delta
_{p}$) & -0.339(31) & \cite{Grossman1999} \\
& 212 & 5 & 210 & 6 & 7s $^{2}$S$_{1/2}$ - 7p $^{2}$P$_{1/2}$ ($\Delta
_{s}-\Delta _{p}$) & 0.009(32) & \cite{Zhang2015} \\
& 212 & 5 & 210 & 6 & 7s $^{2}$S$_{1/2}$ - 7p $^{2}$P$_{1/2}$ ($\Delta _{s}-\Delta
_{p}$) & 0.007(28) & \cite{Grossman1999} \\
& 212 & 5 & 211 & 9/2 & 7s $^{2}$S$_{1/2}$ - 7p $^{2}$P$_{1/2}$ ($\Delta
_{s}-\Delta _{p}$) & -0.334(31) & \cite{Zhang2015} \\
& 212 & 5 & 211 & 9/2 & 7s $^{2}$S$_{1/2}$ - 7p $^{2}$P$_{1/2}$ ($\Delta _{s}-\Delta
_{p}$) & -0.331(34) & \cite{Grossman1999} \\
& 212 & 5 & 213 & 9/2 & 7s $^{2}$S$_{1/2}$ - 7p $^{2}$P$_{1/2}$ ($\Delta_{s}-\Delta _{p}$) & -0.328(34) & \cite{Zhang2015} \\
& 212 & 5 & 221 & 5/2 & 7s $^{2}$S$_{1/2}$ - 7p $^{2}$P$_{1/2}$ ($\Delta_{s}-\Delta _{p}$) & -0.704(42) & \cite{Zhang2015} \\
\\
Ra & 211 & 5/2 & 213 & 1/2 & 7s $^{2}$S$_{1/2}$ - 7p $^{2}$P$_{1/2}$ ($\Delta_{s}-\Delta _{p}$) & 0.6(2) & \cite{Ahmad1983} \\
   & 221 & 5/2 & 213 & 1/2 & 7s $^{2}$S$_{1/2}$ - 7p $^{2}$P$_{1/2}$ ($\Delta_{s}-\Delta _{p}$) & -0.3(8) & \cite{Ahmad1983} \\
   & 223 & 3/2 & 213 & 1/2 & 7s $^{2}$S$_{1/2}$ - 7p $^{2}$P$_{1/2}$ ($\Delta_{s}-\Delta _{p}$) & 0.6(5) & \cite{Ahmad1983} \\
   & 225 & 1/2 & 213 & 1/2 & 7s $^{2}$S$_{1/2}$ - 7p $^{2}$P$_{1/2}$ ($\Delta_{s}-\Delta _{p}$) & 0.4(3) & \cite{Ahmad1983} \\
   & 225 & 1/2 & 213 & 1/2 & 7s $^{2}$S$_{1/2}$ & 0.8(4) & \cite{Arnold1987} \\
   & 227 & 3/2 & 213 & 1/2 & 7s $^{2}$S$_{1/2}$ - 7p $^{2}$P$_{1/2}$ ($\Delta_{s}-\Delta _{p}$) & 0.3(4) & \cite{Ahmad1983} \\
   & 229 & 5/2 & 213 & 1/2 & 7s $^{2}$S$_{1/2}$ - 7p $^{2}$P$_{1/2}$ ($\Delta_{s}-\Delta _{p}$) & 0.6(4) & \cite{Ahmad1983} \\
\\
U & 233 & 5/2 & 235 & 7/2 & 5f$^{3}$6d7s$^{2}$ $^{5}$L$_{6}$ & 0.84(31) & \cite{Gangrsky1997} \\
&  &&&  & 5f$^{3}$6d7s7p $^{7}$M$_{7}$ & 1.32(31) & \cite{Gangrsky1997} \\
&  &&&  & 5f$^{3}$6d7s7p $^{7}$L$_{6}$ & 1.19(89) & \cite{Gangrsky1997} \\
\end{longtable}
\end{landscape}
\newpage

\renewcommand{\arraystretch}{1.0}

\footnotesize

\newpage
\bibliographystyle{plain} 

\bibliography{hfa-2023}

\end{document}